\documentstyle[aps,eqsecnum,prbbib]{revtex}
\begin{document}
\draft
\title{Angular-dependence of the penetration depth in unconventional
superconductors}
\author{Klaus Halterman\cite{klaus}  and Oriol T. Valls\cite{oriol}}
\address{School of Physics and Astronomy and Minnesota Supercomputer
Institute
\\ University of Minnesota \\
Minneapolis, Minnesota 55455-0149}
\author{Igor \v{Z}uti\'c\cite{igor}}
\address{Department of Physics and Center for Superconductivity
Research\\ University of Maryland\\ College Park,
Maryland 20742}

\date{\today}
\maketitle
\begin{abstract}
We examine the Meissner 
state nonlinear electrodynamic effects on the field and angular
dependence of the low
temperature penetration depth, $\lambda$,
of superconductors in several
kinds of unconventional pairing states, with nodes or deep
minima (``quasinodes'') in the energy gap. Our calculations
are prompted by the fact that, for typical
unconventional superconducting material parameters, the predicted
size of these effects for $\lambda$
exceeds
the available experimental precision for this quantity
by a much larger factor than for others.
We obtain expressions for
the nonlinear component of the penetration depth,  $\Delta\lambda$, 
for different two- and
three- dimensional nodal or quasinodal structures. Each
case has a characteristic signature
as to its dependence on the size and orientation
of the applied magnetic field. This shows that $\Delta\lambda$ measurements
can be used to elucidate  the nodal or quasinodal structure of the 
energy gap.
For  nodal lines  we find that
$\Delta\lambda$ is linear in the applied field, while the dependence
is quadratic for point nodes. For layered materials with
$\rm{YBa_2Cu_3O_{7-\delta}}$
(YBCO) type anisotropy, our results for the angular dependence of $\Delta\lambda$
differ greatly from those for tetragonal materials and
are in agreement with experiment.
For the two- and three- dimensional quasinodal cases, $\Delta\lambda$ is no 
longer
proportional to a power of the field and the field and angular
dependences are not separable, with a 
suppression of the overall signal as the node is  filled in.

\end{abstract}

\section{Introduction}

Until about fifteen years ago, the question of
the determination of pairing states
in superconductors was  one of small
and purely theoretical interest,
since no existing superconductors
were commonly suspected\cite{pwave}  to
be in a state other than the standard $s$-wave. Only liquid ${\rm ^3He}$
was known\cite{3he} to exhibit $p$-wave pairing in
its several superfluid phases. Since then, the situation has
dramatically changed. First, extensive studies in 
high temperature oxide superconductors (HTSC's) have 
led to the widespread belief\cite{htsc1} that in most cases the
order parameter for these materials is at least predominantly $d$-wave,
with lines of nodes in a quasi two-dimensional Fermi surface (FS). 
Whether  these are true nodes or very deep minima is, however, not
really established even in the best studied compounds\cite{tk},
and the situation is less clear in some\cite{oth,ncco} other cases.
More recently, unconventional pairing states have been
proposed, on evidence of varying strength, for a plethora
of other materials with lower superconducting
transition temperatures. Among these materials are some
heavy fermion (HF) compounds\cite{hf0,hf,hf2,hf3}, 
members\cite{kan96,organic2,torganic1,organic1,car,ish}
of certain superconducting families of organic
salts such as ${\rm \kappa-(BEDT-TTF)_2C u (NCS)_2}$ and 
 ${\rm (TMTSF)_2X}$ 
(${\rm X = PF_6,ClO_4}$, etc.), and certain other salts
such as\cite{mae,maz,ruth,ruth2,ruth4}  ${\rm Sr_2RuO_4}$
for which a pairing state similar to that in the
A-phase of ${\rm ^3He}$ has been\cite{sameform,sameform2,maki,rice}
suggested, although recently other states\cite{newm} have
also gained favor.

To probe the {\it bulk} order parameter (OP) it is best to use
experimental techniques that measure properties over a scale of the
penetration depth\cite{hardy}  $\lambda$, which in most materials of interest,
as mentioned above, is much larger than the coherence length $\xi$. The
OP at the surface may differ from that in the bulk, and furthermore,
surface experiments are subject to uncertainties arising from
surface quality and preparation problems. In cases where the OP
leads to an
energy gap which
has nodes (or very deep minima which we denote as ``quasinodes''),
it was pointed out eight years\cite{ys} ago, that 
excitation
of quasiparticles  near the nodes by an applied magnetic field
leads to nonlinear anisotropies\cite{sv,ys1,zv1,dahm,amin} in 
the electromagnetic
properties of the material. It was shown\cite{zv2} that one can
in principle
use these anisotropies to perform ``node spectroscopy'', that is, not
only to detect the existence of nodes (or quasinodes) but to infer
their location on the FS.

Although in the earliest work\cite{ys,sv} anisotropies
in the penetration depth were considered  possible subjects
of experimental investigation, emphasis soon switched to related
quantities, chiefly the anisotropic component of the magnetic
moment {\it transverse} to the applied field, and the
torque associated with it. This quantity seemed more accessible
experimentally than $\lambda$, which was deemed to be more
difficult to measure with the requisite precision. However, recent
advances and refinements in experimental techniques force a reconsideration
of this assessment. The best measurements\cite{bhat} of the 
transverse magnetic moment yield only a relatively weak lower bound
on the magnitude of the gap at a quasinode, because the noise
of the measurements is relatively high, with a resulting
uncertainty a factor of only three below the maximum signal expected
for a system with pure nodes. On the other hand, measurements of the
penetration depth in compounds such as YBCO can now be performed\cite{bid}
with a precision of a small fraction of an {\AA}ngstrom. As we show later,
this is one or two orders of magnitude below the putative signal
for that compound.  Furthermore\cite{hv}, 
for many of the 
non-HTSC compounds mentioned above, the predicted
nonlinear signal can be considerably larger than that expected for
HTSC's. Hence, extension of techniques such as those of Ref.\onlinecite{bid}
to dilution refrigeration temperatures is likely to allow
the determination of the nodal structure of these materials.
It appears, therefore, that measurements of
$\lambda$ are the most promising way of probing the nodal structure
of the gap through nonlinear electromagnetic effects.

With this in mind, we discuss here
the nodal spectroscopy
of the penetration depth. We define the quantity 
$\Delta\lambda(\psi) \equiv \lambda(\psi,H_A)-\lambda(\psi,0)$
where $H_A$ is the magnitude of the applied field and $\psi$ the
angle it forms with a suitably defined axis. In defining
$\lambda(\psi,H_A)$
attention must be paid to the experimental methods involved. We
then present results that show that the field and angular
dependence of $\Delta\lambda(\psi)$  will reflect the nodal or quasinodal
structure of the energy gap. We first consider 
quasi two-dimensional sytems with $d$-wave pairing. In this
case, we show that the orthorhombicity 
as it occurs in compounds of the YBCO type,
must be included in a proper
and clear way. We calculate the necessary fields
and find that the effects of orthorhombicity
in $\Delta\lambda(\psi)$
are very important, and that their neglect has
led to misleading conclusions in interpreting experiments.
We then turn to other quasi two-dimensional nodal
and quasinodal cases, for
which the required results are readily obtained from previously 
published field distributions.\cite{zv2} Finally, we consider three
dimensional systems with point or line nodal\cite{hv} structures
with quasinodal admixtures. In our conclusions, we  point
out that the penetration depth can indeed be used to
perform node spectroscopy in all these cases  and elaborate on
the use of our results to interpret existing or future
experiments.

\section{Methods}
\label{methods}

\subsection{Penetration depth}

As explained above, our focus is
on the angular dependence of the penetration depth $\lambda$, as a very 
powerful probe in 
the understanding of the symmetry of the bulk OP
for unconventional superconductors. The 
presence of nodes or quasinodes in the energy gap gives rise\cite{ys,zv2}
to nonlinear corrections in the current response to an applied magnetic 
field, ${\bf H}_{\rm{A}}$. 
These nonlinear corrections result in $\lambda$
having an angular and field dependence 
that reflects directly the symmetry of the pairing state.

The first question we must address is that of defining the angular
dependent penetration depth in the nonlinear case. This involves both
theoretical and experimental difficulties. Let us consider
the geometry of a semi-infinite superconducting slab of thickness $d$,
much larger than any relevant penetration depth. The slab is oriented
perpendicular to one of the symmetry axes, e.g., the $c$ axis, and 
assuming orthorhombic or
higher symmetry, its surfaces are parallel to the plane spanned by
the other two (e.g. the $a$ and $b$) axes. 
This is the geometry that we will
consider in this work. 
In the linear case, the penetration depth
is described in terms\cite{od} of the superfluid density
tensor, and its principal axes are those of symmetry. The 
principal values $\lambda_a$ and $\lambda_b$ can be determined by
experiments involving an
applied field along the $b$ and $a$ directions, respectively. 
Thus one can use, for example, as
a conventional \cite{tinkham} definition,
$\lambda_i \equiv \int_0^{\infty}{\rm d}zH_j(z)/H(0)$, 
where
$H_j(z)$ is the magnetic field along a principal axis, which depends
exponentially on 
the distance $z$ within the sample through only
one of the principal
values. Alternatively, one can write $\lambda_i$
in terms of the spatial derivatives of the fields at the surface, since
both the spatial extent to which $H_A$  
penetrates into the superconductor and its  derivative
at the surface are determined by the same length.
Because of the tensorial nature of the penetration depth, once
the principal values are determined, the results of any experiment
involving applying ${\bf H}_{\rm{A}}$ along an arbitrary  angle $\psi$ with 
respect
to e.g., the $a$ axis can be elucidated, although the result will depend on
the specific experiment considered.

In the presence of nonlinear effects, the situation is much more complicated.
Because of the nonlinearities, the superfluid density or the penetration
length are no longer tensors. Also\cite{zv1,hv} the lengths that 
characterize the
surface derivatives and the extent of field penetration differ by significant
numerical factors. 
Several possible definitions of the effective $\lambda(\psi)$ 
which coincide in the linear 
limit give different results for $\lambda (\psi,H_A)$
in the nonlinear case. 
In general, experimentally one measures the extent to which fields
penetrate and definitions involving  surface derivatives
are not appropriate.
To find the
right definition one must consider the experimental setup.
In the experiments of Ref.~\onlinecite{bid},
the crystal is rotated to different orientations with respect to
the field and a measurement of the component of the magnetic
moment ${\bf m}$ along the applied field is performed. The penetration depth
$\lambda(\psi)$ is then extracted\cite{bidprivate,fw} through the relation:
\begin{equation}
m^{\parallel}(\psi)=-\frac{H_A V}{4 \pi}\biggl(1-\frac{2\lambda(\psi)}{d}\biggr)
\label{pen}
\end{equation}
where $m^{\parallel}$ is the component of ${\bf m}$ along the field and $V$
the volume of the sample. The term in \ref{pen} involving
the effective
penetration depth depends only on the sample area $A$ in the direction
parallel to the field. 
One can equivalently write this
definition of  $\lambda(\psi,H_A)$ in terms of
the integral of the appropriate component of the field
by making use of standard identities.\cite{zv2,jackson,zv3} One then has
\begin{equation}
\label{pen1}
\lambda(\psi,H_A)=\int_{0}^{d/2} dz\, H^{\parallel}(\psi,H_A,z)/H(0),
\end{equation}
where $H^{\parallel}$ is the 
component of ${\bf H}(z)$ parallel to the applied field.
This is the definition of $\lambda(\psi)$ we will use. Other definitions
may have to be employed for different experimental setups. However,
once the nonlinear field distributions inside the
sample are known it is a rather easy matter, as seen below,
to extract 
the effective $\lambda$ corresponding to any other
alternative definition.

To separate the nonlinear effects we write:
\begin{equation} 
\lambda(\psi,H_A) = \lambda_{lin}(\psi) + \Delta\lambda(\psi,H_A)
\label{def}
\end{equation}
Since the linear part is field independent while the nonlinear
part, as we shall see below, vanishes at
zero field, one has that the nonlinear part
$\Delta\lambda(\psi,H_A) = \lambda(\psi,H_A)-\lambda(\psi,H_A=0)$.
We will see that $\lambda_{lin}$ has the expected
angular behavior. The nonlinear part can be written 
in terms of the nonlinear magnetic moment $m_{nl}$ as:
\begin{equation}
\label{nonlinM}
\Delta\lambda(\psi,H_A)
 = \frac{2 \pi}{H_A A} m^{\parallel}_{nl}(\psi,H_A),
\end{equation}
where $m^{\parallel}_{nl}$ is the parallel component of $m_{nl}$.
Since $m_{nl}$ is an extensive quantity, proportional to the sample area, 
we see from (\ref{nonlinM}) that $\Delta\lambda$ is intensive.
Using the expression for the
magnetic moment in terms of  the current field $\bf{j}$, 
${\bf{m}}=\frac{1}{2c}\int{\bf dr (r}\times {\bf j})$, 
standard identities\cite{jackson,zv3}
and the London and Maxwell equations, (\ref{nonlinM}) can be expressed 
entirely\cite{zv2}
in terms of the
value of the nonlinear flow field $\bf{v}$ at the surface of the sample.
This will be done explicitly for several cases. 
We first however, give a brief outline below on the procedure for calculating 
the nonlinear fields in the cases where they are not yet known.

\subsection{Calculation of the fields}
In some of the cases of interest we will be able to use
the field distributions found in previous work\cite{zv2},
but in several others the fields must be calculated.
We explain here briefly the method involved, with details of the
calculations in the Appendices.

Within the framework\cite{ys} of the nonlinear Meissner effect, 
the relation\cite{zv1} between ${\bf j}$ and ${\bf v}$  
is a sum or linear and nonlinear parts, 
${\bf j}({\bf v}) = {\bf j}_{lin}({\bf v}) + 
{\bf j}_{nl}({\bf v})$, where $\bf{v}$ 
is the flow field, and $\bf{j}$ is the supercurrent.
The linear part 
is the usual relation 
${\bf j}_{lin}=-e \tilde{\rho} {\bf v}$, where 
$\tilde{\rho}$ is the superfluid 
density tensor, while the nonlinear term\cite{ys1,zv1} in the low
temperature limit considered here is: 
\begin{equation} 
{\bf j}_{nl}{\bf (v)}= -2 e N^*_f\int_{FS} \, d^2s \: n(s) {\bf v}_f   
\sqrt{({\bf v}_{f} \cdot {\bf v})^2-\left| \Delta(s) \right|^2}  
\Theta(-{\bf v}_{f} \cdot {\bf v}-\left| \Delta(s) \right| ) 
\label{jq}. 
\end{equation} 
Here $N^*_f$ is the total density of states at the Fermi level, 
and $n(s)$ is the 
local density of states 
at the point $s$ on the Fermi surface (FS),  
normalized to unity.
The step function in (\ref{jq})  
restricts the integration over the FS by 
\begin{equation}
\left| \Delta(s) \right|  +{\bf v}_f \cdot {\bf v} <0. 
\label{res}
\end{equation}
The functional relationship between ${\bf j}$ and ${\bf v}$ is then 
combined with  
the Maxwell-London equation,\cite{ys,sv}
\begin{equation} 
\nabla\times\nabla\times{\bf v}=\frac{4 \pi e}{c^2}{\bf j(v)}.
\label{ML}
\end{equation}
Using the boundary conditions,
$\nabla \times {\bf v}|_{d/2}= ({e}/{c}){\bf H}_{\rm{A}}$, 
and  
${\bf {v}}(0)=0$,
we can, for the geometry under consideration,
solve (\ref{ML}) analytically for the necessary fields and then
extract $\Delta\lambda$ from  (\ref{nonlinM}).

\section{Results}
It is convenient to introduce
several dimensionless quantities used in the
calculations. 
These are the dimensionless coordinate 
$\zeta_i$
which represents the coordinate perpendicular to the plane surface of 
the sample
(measured from its midpoint) in units of $\lambda_i$,
the appropriate
penetration depth tensor component in the sample plane.
Its surface
value is 
$\zeta_{s,i} \equiv d/(2\lambda_i)$. The dimensionless 
nonlinear flow field,
\begin{equation}
u_{nl,i} \equiv \frac{v_{nl,i}}{\tilde{v}},
\end{equation}
is normalized by  
the characteristic linear velocity,
\begin{equation}
\label{vchar}
\tilde{v} \equiv \frac{e}{c}\lambda H_A, 
\end{equation}
where $\lambda$ is defined in each case from the in-plane
components of the linear penetration depth.

\subsection{2-D nodal lines with YBCO type othorhombicity}
\label{sybco}
As our first example, we 
examine the nonlinear effects associated with a
two-dimensional gap that has
nodal lines, with crystal orthorhombicity of the YBCO type,\cite{zv2,car99}
that is, with the nonequivalent $a$ and $b$ axes being along the
antinodal directions.
The applied field is in the $a-b$ plane, forming
an angle $\psi$ with the $a$ axis, so that  
the  fields have $a$ and $b$ components, which depend    
only on the coordinate $z$. Considering the usual
linear term only, it is completely elementary to verify that
the definition (\ref{pen}) yields $\lambda_{lin}=
\lambda_b\cos^2\psi+\lambda_a\sin^2\psi$, independent
of the magnitude $H_A$. We can turn then to the nonlinear
$\Delta\lambda$.

The four line nodes are symmetrically placed 
at angles $\varphi_n$ (measured from the positive $a$-axis),
where $n=1,$...,$4$ labels the node.
The Fermi velocity at $\varphi_1$ forms an angle $\alpha$ with 
the $+a$ axis. These angles are shown in Fig.~\ref{FS}.
In the presence of orthorhombic distortion, 
$\varphi_1$  need not equal $\pi/4$, and $\alpha$ does not have
to be equal to $\varphi_1$. One often characterizes the deviation
of $\varphi_1$ from $\pi/4$ by describing the order parameter as
being ``$d+s$'' writing for example $\Delta(\varphi)=\Delta_d \cos(2 \varphi) +
\Delta_s$, and then introducing a 
separate angular variable (see Ref.~\onlinecite{zv2}), 
for the orientation of the Fermi velocity.
This may be misleading, however, since it is in this case no longer
accurate to classify the OP in terms of angular momentum waves. More
important,
since the nonlinear results depend only on the {\it local}
properties at the nodal positions, there is {\it only one}
relevant angular variable, which is
$\alpha$, regardless\cite{old} of which origin one wishes to ascribe
to it.\cite{phot}

We can see from the Fig.~\ref{FS} that 
the magnitude of ${\bf{v}}_f$ is the same at all nodes,
and that we can restrict 
the angle ${\bf H}_{\rm{A}}$ 
makes with the $a$ axis to $\psi \in [0, \pi/2]$.
We characterize the 
anisotropy of the linear
penetration depth tensor
by $\Lambda_a \equiv \lambda_a/\lambda_b$, and it suffices
to take $\Lambda_a \ge 1.$
Only the local properties of the OP near the 
nodes contribute
to the nonlinear current, 
hence we express the OP near the nodes as,
\begin{equation} 
\Delta(\varphi) \approx 2\Delta_0(\varphi-\varphi_n),
\label{ybco} 
\end{equation}
where $\Delta_0$ is half of the slope of the OP near the nodes, and should 
not  necessarily be identified with the gap maximum.

When $\psi \in [0, \pi/2]$, we have (see \ref{res})
the possibility of quasiparticle 
activation (QPA) at the nodes labeled (see Fig.~\ref{FS})
$(1)$ and $(2)$, {\it or} at $(2)$ and $(3)$. 
The specific nodal pair
that is activated depends on which nodal ${\bf v}_f$ satisfy the restriction in 
(\ref{res}).
Anisotropy in the penetration depth tensor
leads to  $\bf v$ twisting (with increased depth  from the surface) 
towards the axis 
with the larger penetration depth, which is $a$. This effect
occurs at linear order and it is therefore very
significant. The nodes which
contribute to the nonlinear term, therefore, depend on the 
dimensionless coordinate $\zeta_i \equiv z/\lambda_i$  
within the sample.
There are three cases to consider:
The first  is when $\psi \in [0, \psi_1]$,
where $\psi_1$ is the
maximum $\psi$ that will result in QPA solely at $(1)$ and $(2)$.
This angle is very small except when $\Lambda_a \gtrsim 1$.
The second case is when $\psi \in [\psi_1, \psi_2]$, 
where $\psi_2$ is the maximum $\psi$ that
will give 
QPA at nodes $(1)$ and $(2)$ until a depth $\zeta_a^*$
is reached, then there is a crossover, and subsequent QPA at nodes 
$(2)$ and $(3)$.
The third region has $\psi \in [\psi_2, \pi/2]$, where
the
only nodes activated at any depth are at $(2)$ and $(3)$.

The nonlinear current-flow relation for 
the case when there is QPA at nodes $(1)$ and $(2)$
is calculated by inserting the OP, Eq.~(\ref{ybco})
into (\ref{jq}). The main steps are carried
out in Appendix \ref{appa}. We find in Eq.~(\ref{acurrent})
for the $a$-component, 
\begin{equation}
\label{jnlatext}
j_{nl,a} = -2\,e\rho \frac{v_a v_b}{v_c}\cos^2\alpha \sin\alpha.
\end{equation}
Here $\rho \equiv \frac{1}{2}N_f v_f^2$, is the local value of
the superfluid density at the nodes, 
$N_f$ is the local density of states,
$v_f$ is the Fermi speed  at the nodes, and 
the local critical velocity is
$v_c \equiv \Delta_0/v_f$. 
Similarly,  when there is QPA at $(2)$ and $(3)$,
the $a$-component of the current is given by (\ref{acurrent2}),
\begin{equation}
\label{jnlatext2}
j_{nl,a} = \frac{e\rho}{v_c} \cos\alpha\left[v_a^2 \cos^2\alpha + 
v_b^2 \sin^2\alpha\right].
\end{equation}
Analogous expressions can be written for the $b$-component.
To find the nonlinear flow field 
we insert the 
current
(\ref{jnlatext}) into (\ref{ML}), for 
$\zeta_a \in [\zeta_a^*,\,\zeta_{s,a}]$
or  (\ref{jnlatext2}) for
$\zeta_a \in [0,\,\zeta_a^*]$. 
The solution is found perturbatively (to first order)
using the previously stated boundary
conditions plus continuity of the flow field, magnetic field, and current at
the crossover point $\zeta_a^*$. The details are given in  Appendix \ref{appb},
the main results are (\ref{unla}) and (\ref{unlb}).
We choose $\lambda \equiv \sqrt{\lambda_a \lambda_b}$ as the normalization in
(\ref{vchar}), and
we take\cite{zv2} $\lambda=\lambda_n$, its nodal value.
$H_0$ is the usual\cite{zv1,zv2}
characteristic field,
\begin{equation}
\label{H_0}
H_0 = \frac{c \Delta_0}{e \lambda v_f}.
\end{equation}

We can now achieve our objective, and get $\Delta\lambda$ from the 
calculated fields. We 
write
(\ref{nonlinM}) in terms of the dimensionless flow field,
\begin{equation}
\label{lambdaortho}
{\Delta\lambda} = 
\lambda\left[\sin\psi \,u_{nl,a}(\zeta_{s,a}) - \cos\psi 
\,u_{nl,b}(\zeta_{s,b})\right].
\end{equation}
The flow field results in  Appendix \ref{appb}
are valid for any material thickness $d$. Upon taking
the slab limit $d \gg \lambda$, we can express
$\Delta\lambda(\psi)$ in the following form:
\begin{equation}
\label{calp}
\Delta\lambda(\psi,H_A) = \frac{1}{6} \frac{H_A}{H_0} \lambda {\cal Y}(\psi).
\end{equation}
We see from this result that the
nonlinear effect in the penetration depth is
proportional to the field, for line nodes. The quantity
${\cal{Y}}(\psi)$ represents its angular dependence,
normalized so that its maximum is unity for the tetragonal case.
The function $\cal{Y}$ has a different expression for each region of 
$\psi$, with the crossover angles $\psi_1, \psi_2$ being given in 
(\ref{xangles}). These expressions are:
\begin{mathletters}
\begin{eqnarray}
\label{ybcowhole}
{\cal{Y}}(\psi) &=& \frac{18 \Lambda_a}{2+\Lambda_a}\cos^2\alpha\sin\alpha\sin^2\psi\cos\psi+
\frac{2}{\Lambda_a^2}\sin^3\alpha\cos^3\psi, \qquad \psi\in [0,\psi_1], \\ 
{\cal{Y}}(\psi) &=& \frac{18 \Lambda_a}{2+\Lambda_a}\cos^2\alpha\sin\alpha\cos\psi\sin^2\psi \\ \nonumber
&+&\frac{2}{\Lambda_a^2(1+2\Lambda_a)}\sin^3\alpha\cos^3\psi\Bigl[1+2\Lambda_a+(4\Lambda_a-1)
\left(\frac{\Lambda_a \tan\psi}{\tan\alpha}\right)^{\frac{3\Lambda_a}{\Lambda_a-1}}\,\Bigr] \\ \nonumber
&+& \frac{2\Lambda_a^2(2\Lambda_a^2-10\Lambda_a-1)}{(2+\Lambda_a)(1+2\Lambda_a)}\cos^3\alpha\sin^3\psi
\left(\frac{\Lambda_a\tan\psi}{\tan\alpha}\right)^{\frac{3}{\Lambda_a -1}}, \qquad \psi\in [\psi_1,\psi_2],\\
{\cal{Y}}(\psi) &=& \frac{18}{1+2\Lambda_a}\sin^2\alpha\cos\alpha\cos^2\psi\sin\psi+
{2}\Lambda_a^2 \cos^3\alpha\sin^3\psi, \qquad \psi\in [\psi_2,\frac{\pi}{2}].
\label{ybcowhole3}
\end{eqnarray}
\end{mathletters}
When $\Lambda_a=1$, $\psi_1=\psi_2=\alpha$ and the middle one 
of the 
expressions above is not needed.
We present plots
for  ${\cal{Y}}(\psi)$ in the next three Figures, 
where $\psi$ is limited to  $0<\psi<\pi/2$,
since the result in the
remaining range is trivially obtained by symmetry.
In Fig.~\ref{varyL},
we show results for fixed
$\alpha = \pi/4$, and vary the anisotropy
parameter  $\Lambda_a$ of the penetration depth tensor,
which influences the result through the twisting  of
the fields inside the sample. For reference, 
we  plot
the isotropic result as the bold curve. The ratio
of ${\sqrt{2}}/{2}$ between the minima and maxima of this reference
curve has been known for a long time.\cite{ys}
We consider the range $1.0-1.5$ for
$\Lambda_a$, in increments of 0.1. We see that as this parameter
increases, the symmetry of the curve changes, to reflect the mixing
of ${\pi}/{2}$ and $\pi$ symmetries. 
The signal increase with $\Lambda_a$  when $\psi={\pi}/{2}$, can readily
be understood physically, since the volume occupied by the currents
will be in this case determined by the larger of the linear components.
We can see from the Figure 
the signal characteristics change considerably with increasing anisotropy.
The maxima and minima of $\cal{Y}$ are no
longer separated by the factor of ${\sqrt{2}}/{2}$ for $\Lambda_a>1.0$.
With sufficiently large $\Lambda_a$, as seen in the Figure,
the signal  at $\psi={\pi}/{4}$ becomes approximately
the average
of ${\cal{Y}}(\psi=0)$ and ${\cal{Y}}(\psi={\pi}/{2})$. 
This illustrates the high sensitivity
of the penetration depth
to anisotropy, as compared with 
the previously studied\cite{zv2} transverse
magnetic moment. The reason for this difference in the transverse 
and longitudinal
behaviors is explained in the last Section.

Next, in Figure~\ref{alpha}, we examine the effects of varying $\alpha$ while 
keeping $\Lambda_a =1.0$ fixed.
Again, the bold curve is the $\alpha = {\pi}/{4}$ result, and the other 
curves are for values $\alpha=(\pi/4) \pm n\delta \alpha$ with $n=1, 2, 3$ and
$\delta\alpha={\pi}/{80}$.
It is seen that, at $\psi = 0$, the signal increases 
as $\alpha$ increases above $\pi/4$, reflecting the increase in 
$|{\bf v} \cdot {\bf v}_f|$. For smaller
values of $\alpha$ the effect reverses, with the curves
corresponding to the same $n$ and opposite sign
being symmetric with respect to exchange of the $a$ and $b$
axes. This behavior
is very sensitive 
to small changes in $\alpha$.
The effects of increasing $\alpha$ are in a sense opposite to those
of increasing $\Lambda_a$.
It is interesting therefore to see how these two effects combine.
Thus, in Fig.~\ref{both}, we  plot the normalized $\Delta\lambda$ 
for $\Lambda_a =1.3$, and the same
values of $\alpha$ as in Fig.~\ref{alpha}. The symmetry of
the curves for the same $n$ is now lost, because
$\Lambda_a \ne 1$. The overall
conclusion that one can draw from these results is that in the presence
of even relatively moderate deviations from the tetragonal
``pure $d$-wave'' situation, the appearance of the $\Delta\lambda(\psi)$
vs $\psi$ data might reflect more a $\pi$ than a ${\pi}/{2}$ symmetry.
This has to be kept present in analyzing experimental results.
We will return to this point in Section \ref{conclusions}.

\subsection{2-D line nodes/orthorhombicity of BSCCO type}
\label{sbssco}
Next we consider the nonlinear effects associated with 
a two-dimensional OP with line nodes in the presence of orthorhombic
anisotropy characteristic of $\rm{Bi_2Sr_2CaCu_2O_{8+\delta}}$ (BSCCO) materials. The new
orthorhombic $a$ and $b$ axes of symmetry form angles
of ${\pi}/{4}$ with the undistorted tetragonal axes.
In this case, 
the nodes are at angles of ${\pi}/{2}$ from each other and the nodal
Fermi velocities are aligned with the nodal directions, which
are the  principal axes of the system. However, 
the  two nodes on the new $a$ axis
are not equivalent to the other two
on the new $b$ axis. The fields
for this case were calculated in Section IIIA.2 of Ref.~\onlinecite{zv2}.
We need only to insert these fields
into (\ref{lambdaortho}) and get (for $d\gg\lambda_i$),
\begin{equation}
\Delta\lambda(\psi,H_A)=\frac{1}{6}\frac{H_A}{H_0}\lambda {\cal{B}}(\psi), 
\end{equation}
where $\lambda \equiv \sqrt{\lambda_{na}\lambda_{nb}}$, 
with $\lambda_{ni} \equiv ({2 \pi e^2}/{c^2})N_{f}
v_{fni}^2$, where
$v_{fni}$ are the nodal Fermi velocities, $H_0$ is defined in 
(\ref{H_0}) with $v_f = \sqrt{v_{fna} v_{fnb}}$, and we take
$\lambda_{ni} = \sqrt{\lambda_a\lambda_b}$.
Here the simple angular dependence is contained in the factor ${\cal{B}}(\psi)$, 
and is given by
\begin{equation}
\label{2dbssco}
{\cal{B}}(\psi)=\Lambda_a^{-1/2} \cos^3\psi + \Lambda_a^{1/2} \sin^3\psi.
\end{equation}
In Fig.~\ref{bsscofig}, we show ${\cal{B}}(\psi)$ for varying degrees of 
anisotropy, $\Lambda_a =
1.0-1.5$ in increments of $1/10$.

\subsection{2-D quasinodal lines}
\label{s2dqn}
Now we turn to the situation
where there are no nodes, but rather,
very deep minima in the gap function (quasinodes). This
can be due to a small, constant 
$i\Delta_s$ or $i\Delta_{d_{xy}}$
component in the OP, which we assume is added to a main
$\Delta_{d_{x^2-y^2}}$  component. The OP near the quasinodes can
then be written in the form
\begin{equation}
\Delta(\varphi) = i\Delta_{\rm {min}} + 2\Delta_0(\varphi-\varphi_n),
\end{equation}
where $2 \Delta_0$ is the slope of the OP at the minima of the gap 
function, and
$\Delta_{\rm {min}} \ll \Delta_0$ is the minimum value of the energy gap.
Here we consider only the isotropic case $\lambda_a=\lambda_b \equiv \lambda$, 
$\alpha={\pi}/{4}$, 
which is sufficient to illustrate the changes brought about by the presence
of quasinodes, rather than nodes, and 
where we can use previously calculated field distributions from 
Section IV of Ref.\onlinecite{zv2}.  We insert these fields
into (\ref{pen}). The resulting nonlinear 
$\Delta \lambda$ depends on two variables, besides the angle $\psi$: the
ratio  $h \equiv H_A/H_0$ of the applied field to the characteristic
field $H_0$, defined in (\ref{H_0}), and the ratio 
\begin{equation}
\kappa \equiv \frac{\Delta_{\rm {min}}}{\Delta_0}\biggl/\frac{H_A}{H_0}.
\label{kappaeq}
\end{equation}
We find after straightforward algebra, in the $d \gg \lambda$ limit,
\begin{equation}
\label{2dquasi}
\Delta\lambda(\psi,H_A)= \frac{1}{6}\lambda\frac{H_A}{H_0}{{\cal{Q}}}(\psi,\kappa).
\label{2dqn}
\end{equation}
The angular and field dependences now no longer factorize
and are (apart from the overall factor of $h$) represented
by the function of two variables ${\cal Q}$, which
is normalized to unity at $\psi=0$, $h=0$:
\begin{eqnarray}
\label{Q}
{\cal{Q}}(\psi,\kappa)&=&(\cos\psi^\prime - \kappa)^2 (2\kappa + 
\cos\psi^\prime)\Theta\left(\cos\psi^\prime-\kappa\right) \\ \nonumber
&+&(\sin\psi^\prime-\kappa)^2
(2\kappa+\sin\psi^\prime)\Theta\left(\sin\psi^\prime-\kappa\right),
\end{eqnarray} 
written for $0\le\psi^\prime\le\pi/2$ (trivially
extended by symmetry to the remaining range)
with $\psi^\prime\equiv \psi - \pi/4$.
The step functions imply that the nonlinear effects vanish for
$h <\delta$, with $\delta \equiv \Delta_{\rm {min}}/\Delta_0$, since 
then the field is not sufficiently strong to create
quasiparticles of energy larger than the minimum gap value.
The behavior of ${\cal{Q}}(\psi,\kappa)$ is
plotted 
in Fig.~\ref{quasi2d}, where we show its angular dependence
for several values of $\kappa$, with the applied field above threshold.
One can see how the filling of the node produces a fast decrease
in the nonlinear effect on the penetration depth, as it does\cite{bhat}
also for the transverse moment.
In Fig.~\ref{quasi2dh}, the variation of 
${\Delta\lambda}/{\lambda}$
as a function of $H_A/H_0$ is shown
for several values of $\delta$, at $\psi=\pi/4$. The 
field threshold effect is clearly
seen, and can be read off directly from the curves. 

\subsection{3-D quasinodal points and lines}
\label{3d}
We now examine cases where the FS is three-dimensional and the nodal
or quasinodal structure of the
energy gap involves points or lines. We will consider here the
same situation for which, in the limit of pure nodes,  the transverse
magnetic moment was calculated in Ref.~\onlinecite{hv}. For the
sake of brevity, we will compute directly the fields in the
quasinodal case, and consider the situation where actual
nodes exist as the appropriate special limit.
For both of the cases
considered below, we will assume that the slab surfaces are
parallel to the $a-c$ plane, and that
${\bf H}_{\rm{A}}$ is also in this plane, with $\psi$ defined with
respect to the $c$ axis.
Then, the nonlinear 
fields have only $x$ or $z$ components (depending on whether we are discussing
lines or point nodes respectively, see below),
which depend only on the coordinate $y$ normal to the
sample. 
Again  it is elementary to
show that $\lambda(\psi,H_A) = \lambda_{lin}(\psi) + \Delta\lambda(\psi,H_A)$, 
where $\lambda_{lin}(\psi) = \lambda_z\sin^2\psi+\lambda_x\cos^2\psi$, 
independent
of $H_A$ and
\begin{equation}
\label{pen3d}
\Delta\lambda(\psi,H_A) =\lambda_z\sin\psi\, u_{nl,z}(\zeta_{s,z})
-\lambda_x\cos\psi\, u_{nl,x}(\zeta_{s,x}).
\end{equation}
We will use this result with the fields calculated below.
\subsubsection{3-D point nodes}

We first consider an OP 
leading to a gap with two quasinodes at the poles along the $z$-axis.
In this configuration,  
the nonlinear fields have only a $z$-component, which depends    
on the coordinate $y$. By symmetry, we can restrict our analysis to the 
quasinode at $\theta=0$,
where $\theta$ is the usual polar angle. 
We take the form of the gap near the quasinode to be
\begin{equation}
|\Delta(\theta)| = (|\Delta_{\rm {min}}|^2 + |\Delta_p\theta|^2)^{1/2},
\qquad \theta \approx 0,
\label{ps}
\end{equation}
where $\Delta_p$ is the slope of the OP near the node, 
and $\delta \equiv \Delta_{\rm {min}}/\Delta_p \ll 1$. This is the generalization
of the previously studied\cite{hv} OP to the quasinodal case.

The nonlinear current response as
a function of the flow field is calculated in  Appendix \ref{appa}, and we find,
for its only nonzero component,
\begin{equation}
\label{3dpointj} 
j_{nl,z}=\frac{e\rho}{v_p^2 }\, \left(v_z^2-v_{s}^2 \right)^{3/2}\Theta(v_z-v_{s}),
\end{equation}
where $v_{p} \equiv \Delta_{p}/v_f$, $v_s \equiv \Delta_{\rm{min}}/v_f$,
 and in three dimensions, $\rho \equiv 
\frac{1}{3}N_f v_f^2$ in terms of local values.
This result shows that
there are no nonlinear effects present if $v_z < v_{s}$.
Since the flow field decreases  with distance into the sample, there
will be a depth in the material, which, in terms
of the dimensionless variable $\zeta_z \equiv y/\lambda_z$,
we denote as $\zeta_z^*$, where $v_z=v_{s}$, so that
nonlinear corrections are absent for the region below $\zeta_z^*$.
Inserting (\ref{3dpointj}) into (\ref{ML}), we get an equation for
the flow field. This equation
can be solved perturbatively to first order in the small parameter 
$(H_A/H_0)^2$, where $H_0 \equiv {c \Delta_p}/{e \lambda_z v_f}$. 
In Appendix \ref{appb} we find, taking
into account these subtleties, the solution $u_{nl,z}(\zeta_z)$ for arbitrary 
thickness $d$. The result is given in (\ref{3dsol}).
We then have from (\ref{pen3d}) and (\ref{3dsol}),
after taking the limit $d\gg\lambda_z$, the result, valid for $0<\psi<\pi/2$,
\begin{equation}
\label{lampoints}
\Delta\lambda(\psi,H_A) = \frac{1}{4}\lambda_z \frac{H_A^2}{H_0^2}
{\cal{P}}(\psi,\kappa)
\Theta(\sin\psi-\kappa),
\end{equation}
where the function ${\cal{P}}(\psi,\kappa)$ is given by,
\begin{equation}
{\cal{P}}(\psi,\kappa)=\frac{3}{2} \kappa^4\ln\left(\frac{\sin\psi + \sqrt{\sin^2\psi-\kappa^2}}{\kappa}
\right)+\left(\sin^3\psi - \frac{5}{2} \kappa^2 \sin \psi\right)\sqrt{\sin^2\psi - \kappa^2}.
\end{equation}
Here we see the mixed angular and field dependence of the result, 
as in the
function ${\cal Q}$ in (\ref{Q}).
The step function parameter, 
involving
the quantity
\begin{equation}
\kappa \equiv \frac{\Delta_{\rm {min}}}{\Delta_p}\biggl/\frac{H_A}{H_0},
\label{kappa2}
\end{equation}
reflects again the requirement that the appropriate
field component exceed
a minimum value.
In Fig.~\ref{3dpoints}, we plot ${\cal{P}}(\psi,\kappa)$, for various values of 
$\kappa$ with the applied field above threshold. One again can see the
decrease of the effect as the minimum gap value increases.
The inset contains the
field dependence of ${\Delta\lambda}/{\lambda_z}$
at $\psi=\pi/2$, for $\delta = 0 - 0.04$, 
in increments of 0.01. There we can see that the field dependence 
above threshold is no longer parabolic,
as is the case when $\Delta_{\rm {min}} = 0$.

\subsubsection{3-D line nodes}
Finally, we consider a three dimensional FS
with a quasinodal line in the $x-y$ plane.
The angular dependence of the gap near the quasinodal line 
at $\theta = {\pi}/{2}$ has the form,
\begin{equation}
|\Delta(\theta)| = (|\Delta_{\rm {min}}|^2 + |\Delta_p(\frac{\pi}{2}-\theta)|^2)^{1/2},
\qquad \theta \approx \frac{\pi}{2},
\label{pcs}
\end{equation}
with  $\Delta_{\rm {min}} \ll \Delta_p$.
The transverse magnetic moment
in the $\Delta_{\rm {min}}=0$ limit of this
OP has been previously\cite{hv}
studied. For this case,
the fields only have $x$ components, and we assume
tetragonal symmetry.
We calculate the nonlinear current in Appendix \ref{appa}
where we find,
\begin{equation}
j_{nl,x} = \frac{e \rho}{v_p v_x} 
\left(v_{x}^2-v_{s}^2\right)^{3/2}\Theta(v_{x}-v_{s}).
\label{explicitjctext} 
\end{equation} 
Again, the nonlinear current vanishes unless the flow field, 
$v$, is sufficiently large; $v_{x}> v_{s}$. 
Once (\ref{explicitjctext}) is inserted into ({\ref{ML}}), we get an 
equation for the
flow field that can be solved with a procedure identical to the point node 
case.
This is done in Appendix \ref{appb}.
We find in the thick slab limit, for the only nonzero component:
\begin{equation}
\label{unlxtext}
u_{nl,x}(\zeta) = \frac{H_A}{H_0}\left[E_1 e^{\zeta} + 
E_2 e^{-\zeta} + r(\zeta)e^{\zeta} + s(\zeta)e^{-\zeta}\right],
\end{equation}
where $H_0 \equiv c\Delta_p/e\lambda v_f$, 
$\zeta \equiv y/\lambda$, $\lambda$ is the linear penetration depth
in the $a-b$ plane,
and the constants $E_1$, $E_2$, and the functions $r$, and $s$ are 
given in Appendix \ref{appb}.
The nonlinear correction 
to the penetration depth is then obtained from (\ref{pen3d})
and (\ref{unlxtext}), as:
\begin{equation}
\label{lamline}
\Delta\lambda = \frac{1}{3} \lambda \frac{H_A}{H_0} {\cal{L}}(\psi,\kappa)
\Theta(\cos \psi-\kappa),
\end{equation}
where ${\cal{L}}(\psi,\kappa)$ is normalized so
that its maximum at $\kappa = 0$ is
unity,
\begin{equation}
{\cal{L}}(\psi,\kappa)=3 \kappa^3 \tan^{-1}\left(\frac{\sqrt{\cos^2\psi-\kappa^2}}{\kappa}\right)+
\left(\cos^2\psi-4\kappa^2\right)\sqrt{\cos^2\psi-\kappa^2}.
\end{equation}
The overall power law behavior at $\Delta_{\rm {min}}=0$ is now linear
in the field, and at finite $\Delta_{\rm {min}}$ a threshold effect is found.
We plot these results in  
 Fig.~\ref{3dline}, where we 
display ${\cal{L}}(\psi,\kappa)$, for various values of
$\kappa$. The resulting
behavior is very reminiscent from that
found in the previous case.
The inset contains the field dependence of ${\Delta\lambda}/{\lambda}$
at $\psi=0$, 
with  $\delta$ having the same values as in Fig.~\ref{3dpoints}.
Because of the mixed dependence
of these results on the field and angular variables, the 
curves shown
are not linear, except in the case $\delta = 0$, and again 
the threshold values can be read
off directly from the intercepts.

\section{Discussion and Conclusions}
\label{conclusions}
In summary, prompted by recent refinements in experimental techniques
that allow very high precision 
measurements of $\Delta\lambda$, and by the ever increasing
number of candidates for unconventional superconducting states,
we have calculated the low temperature angular and field dependence of the 
nonlinear
component to the penetration depth for several different
two and three dimensional energy gaps with nodal or quasinodal structures. 

The expected  signal 
for the nonlinear penetration depth effect exceeds the
available  experimental resolution\cite{bid,gia} of one 
tenth of an $\rm{\AA}$ngstrom by a considerable factor.
To see this, consider for example a tetragonal compound with
pure $d$-wave pairing and a linear penetration depth $\lambda=1400 \rm{\AA}$,
in the  YBCO range. From (\ref{calp}), we easily get that the difference
$\delta\lambda \equiv \Delta\lambda_{\rm{max}}-
\Delta\lambda_{\rm{min}}$,
between the minimum and maximum values of $\lambda(\psi)$ would be,
at $h=0.04$, about 3 $\rm{\AA}$, a factor of thirty better than
the experimental resolution.  
This is an order of magnitude improvement when compared to the corresponding
estimate, under the  same assumptions, for 
measurements of the 
the transverse magnetic moment where one has at best\cite{bhat}
a factor of three.
Further, taking into account the orthorhombicity
of  YBCO by setting\cite{basov} $\Lambda_a = 1.6$, which increases
 $\delta \lambda$
(see Fig.~\ref{varyL}),
we find from the same equation  and at the same $h$, 
$\delta\lambda \approx 
14 \rm{\AA}$. This is a factor of 140 above the resolution  
achieved in Refs.~\onlinecite{bid,gia}.

For gap functions with line nodes, we have found that $\Delta \lambda$ at fixed
angle is proportional to the field. Our results, however, are
obtained in the low temperature, clean limit. Both finite
temperature\cite{ys1} and  impurities\cite{ys1,sv} modify this linear
behavior at smaller fields, where nonlocal effects\cite{hirshy} may have
a similar influence.\cite{com} Since the combined
outcome of these effects is not at present amenable to 
reliable computation,
it is safer to perform the experiments and to compare with theory
at the largest possible fields, where the behavior should approach
linearity. $H_A$ can be increased all the way to the
field of first flux penetration, $H_{f1}$, taking in this full advantage
of this field being in practice\cite{bhat,com}  much larger than 
the Ginzburg-Landau estimate of $H_{c1}$. 

Keeping this in mind,
let us consider our results 
for two-dimensional gap functions with nodes 
for materials with YBCO-type orthorhombicity. 
Some experimental results
for the angular dependence of
$\lambda(\psi)$ 
for YBCO are available,\cite{bid} although only for a few
selected directions. We have found that
the angular dependence of $\lambda(\psi)$ is
extremely sensitive (see Fig.~\ref{varyL}) to
small departures from unity in the anisotropy factor $\Lambda_a$, 
eventually resulting in a change in the apparent
leading symmetry behavior of $\Delta\lambda(\psi)$, which then looks quite
different from that found in the tetragonal case.
Taking again $\Lambda_a=1.6$, for\cite{basov} YBCO,
we find that
${\cal{Y}}(\psi=0)\approx 0.30$,
${\cal{Y}}(\psi=\pi/2)\approx 1.8$,  
while ${\cal{Y}}(\psi=\pi/4)\approx 1.2$, very close to the
average in the two main axial directions. This is precisely what it is
found experimentally.\cite{bid}  Because this is so different
from what happens when $\lambda_a=\lambda_b$, it was mistakenly
interpreted in the experimental work as evidence
against, instead of for, the anisotropy
found there being due to the nonlinear Meissner effect. The measured field
dependence\cite{bid} departs considerably at small fields
from linearity. This is apparently \cite{bid,bidprivate} not due
to temperature effects alone and we believe it is very likely 
attributable to impurities,
since the zero field temperature dependence of $\lambda$ 
of the sample used departs
appreciably from linearity for temperatures below about 3 K. At the
largest fields, the field dependence extrapolates to linear, with
reasonable values of $H_0\approx 9000$ gauss. Thus, these experimental
results are consistent, as far as their field and angular
dependence, with our theory. The weak temperature dependence of these
and other\cite{gia} measurements
remains, however, a puzzle. It cannot
be ruled out, given the complications involving the correct
treatment of impurity averaging\cite{atk} in these materials, that
the temperature has a relatively weak effect in the 
samples studied. Further experimental work in the same or
other materials is needed. Preliminary results
for single crystals of TI-2201
show $\Delta\lambda$ having a linear magnetic field dependence that is
interpreted\cite{boulder} as agreeing with theoretical
expectations for the nonlinear 
Meissner effect.

This strong sensitivity of $\Delta\lambda$
to anisotropy (either to $\Lambda_a$ or to $\alpha$,
the angle that 
${\bf{v}}_f$, at the node, makes with the $+a$ axis)
would not be expected
from previous calculations\cite{zv2} 
of the transverse magnetic moment, where the effects of orthorhombicity were
not pronounced. The reason is that the transverse moment is constrained
by symmetry
to vanish, regardless of orthorhombicity, both at $\psi=0$ and
at $\psi=\pi/2$, plus at one point in between. This constraint
does not exist for a longitudinal measurement.

We have also examined gaps with two-dimensional quasinodes, and found 
that the field and angular 
dependence are no longer
separable. The angular and field dependence  of $\Delta\lambda$ is
governed by a term
linear in the field and by a step
function indicating that a minimum threshold field must
be applied to excite quasiparticles
above the gap minimum. This
is multiplied by a function of $\psi$ and of the parameter 
$\kappa$, which is a ratio (see (\ref{kappaeq})) relating the
value of the gap minimum to the applied field strength.  
The signal decreases markedly as $\kappa$ increases.

In Subsection \ref{3d}, we investigated three dimensional gaps with points 
and line quasinodes.
There again the nonlinear contribution to 
the penetration depth depends
on a function of  angle and of a parameter $\kappa$
now defined in (\ref{kappa2}), a step function,
and a separate factor
linear in the applied field for line nodes and quadratic for points.
The
situation is similar, as far as the field
dependence, to that for the
two-dimensional case. The 
signal decreases
with increasing $\kappa$ and vanishes at threshold. 
For example, at $\kappa = 0.6$, the nonlinear 
signals for both points and lines drop
to about 25\% of their maximum ($\kappa = 0$) values. 
Even with such 
large admixtures, however,
the signal is still likely to be
within current experimental resolution.
Let us
estimate the signal for
an OP with three dimensional line nodes, similar
to that which
might occur in  ${\rm Sr_2RuO_4}$, or certain\cite{hf2}
heavy fermion compounds.
Using (\ref{lamline}), with $\kappa = 0$, we find, $\delta\lambda = 
\Delta\lambda_{\rm {max}}=\frac{1}{3}\lambda h$.
Using published\cite{ruth}
values for  ${\rm Sr_2RuO_4}$ we estimate
$\lambda = 2000 {\rm{\AA}}$ and a value  $h=0.3$ for
$H_A=H_{c1}$. These values give
a maximum signal of,
$\delta\lambda \approx 200 {\rm{\AA}}$. The magnitude of the signal in 
this case is well
above experimental resolution, and even with  relatively large 
admixture leading to a substantial 
$\Delta_{\rm {min}}$ 
the signal would still be experimentally tangible.

We have focused here on the nonlinear effects on the angular 
dependence of the penetration depth, 
and we have shown the strong influence that anisotropy
in  the
principal values of the linear penetration depth and
orientation of
the Fermi velocity has on the results. 
The methods presented in this paper can  be readily extended to 
other nodal patterns and to include
the nonlinearities in the temporal response that arise from a time-dependent 
magnetic field.\cite{zv4,scal} These
phenomena are currently being investigated via
microwave measurements.\cite{anlage}

\acknowledgments
We thank A. Bhattacharya for many conversations, 
C. Bidinosti for providing us with much information about
his experimental methods and S.M. Anlage for insightful
discussions with I.\v{Z}. This work was supported in part
by the Petroleum Research Fund, administered by the ACS (at Minnesota)
and by the US ONR Grant N000140010028 and DARPA (at Maryland).        

\appendix
\section {currents}
\label{appa}

\subsection{2-D nodal lines with orthorhombicity}
For the order parameter given by (\ref{ybco}), the four line 
nodes are symmetrically placed (see Fig.~\ref{FS})
at angles $\varphi_n$, measured from the positive $a$-axis,
where $n=1$,...,$4$ labels the node.
The Fermi velocity 
at node 1 is ${\bf v}^{(1)}_f = v_f(\cos \alpha,\sin \alpha)$. 
After making the replacement  
$N^*_f\int_{FS} d^2s \: n(s)\rightarrow N_f\int_{\bar{\varphi}_c}   
\:  d\varphi /2 \pi$, Eq.~(\ref{jq}) gives 
the contribution, ${\bf j}_{nl}^{(n)}$, when quasiparticles at $\varphi_n$
are activated: 
\begin{equation}
{\bf{j}}^{(n)}_{nl}{\bf (v)}= -2 eN_f\int^{\bar{\varphi}_c}_{-\bar{\varphi}_c} 
 \frac{d\bar{\varphi}}{2\pi} \: 
{\bf{v}}_f\sqrt{(2\Delta_0 \bar{\varphi}_c)^2-( 2\Delta_0 \bar{\varphi})^2}
\label{jqa},
\end{equation}
where $\bar{\varphi} \equiv \varphi-\varphi_n$, 
and the integration is limited by 
$\bar{\varphi}_c = |{{\bf v}_f \cdot {\bf v}}|/{(2\Delta_0) }$.
One finds,
\begin{equation} 
{\bf j}^{(n)}_{nl}=-\frac{e}{4 \Delta_0}N_f{\bf v}^{(n)}_f[{\bf v}^{(n)}_{f} 
\cdot {\bf v}]^2.
\label{jnla1} 
\end{equation}
Except when $\bf v$ is along a nodal Fermi velocity,
in general two nodes must be
considered.
If the nodes at $\varphi_1$ and $\varphi_2$ are activated, 
we can get the total 
nonlinear current by adding ${\bf{j}}^{(1)}_{nl}+{\bf{j}}^{(2)}_{nl}$ from 
(\ref{jnla1}):
\begin{mathletters}
\begin{eqnarray}
\label{acurrent}
j_{nl,a} &=& -2\,e\rho \frac{v_a v_b}{v_c}\cos^2\alpha \sin\alpha,\\
j_{nl,b} &=& -\frac{e\rho}{v_c} \sin\alpha\left[v_a^2 \cos^2\alpha + v_b^2 
\sin^2\alpha\right].
\label{bcurrent}
\end{eqnarray}
\end{mathletters}
where we have introduced
the local superfluid density, $\rho \equiv (1/2)N_f v_f^2$, and 
critical velocity
$v_c = \Delta_0/v_f$.
Likewise, if the nodes at $\varphi_2$ and $\varphi_3$ are activated, we get
\begin{mathletters}
\begin{eqnarray}
\label{acurrent2}
j_{nl,a} &=& \frac{e\rho}{v_c} \cos\alpha\left[v_a^2 \cos^2\alpha + v_b^2 
\sin^2\alpha\right], \\
j_{nl,b} &=& 2\,e\rho \frac{v_a v_b}{v_c}\sin^2\alpha \cos\alpha.
\label{aacurrent}
\end{eqnarray}
\end{mathletters}

\subsection{Nonlinear current for 3-D quasinodes}
\subsubsection{3-D point quasinodes}

We examine first a gap of the form  (\ref{ps}).
By symmetry, we can restrict ourselves to the node at $\theta=0$
since the contribution from $\theta=\pi$ is identical.
Thus, ${\bf v}_f \approx (0,0,v_{fz} )$, and the relevant region of integration
is limited by $\theta_c$, as determined from
$({\bf v}_f \cdot {\bf v})^2=\left| \Delta(\theta_{c}) \right|^2$.
In performing the integral in (\ref{jq}) we again
replace $N_f^*\int_{FS} d^2s \: n(s)$ by 
$N_f\int_{\Omega_c} \: d\varphi \theta d\theta/4 \pi$.
This yields  
only a $z$-component to the nonlinear current, 
\begin{equation}
j_{nl,z}= -\frac{ e N_f v_f \Delta_p}{2\pi} \int^{2\pi}_{0} d\varphi 
\int^{\theta_c}_{0} \theta d\theta 
(\theta_c^2-\theta^2)^{1/2}\Theta(v_z-v_{s}),
\end{equation}
where
$\theta^2_{c}\equiv [(v_fv_z)^2-\Delta_{\rm {min}}^2]/\Delta_p^2$.
We get,
\begin{equation}
j_{z}=\frac{e\rho}{v_p^2 }\, \left(v_z^2-v_{s}^2 \right)^{3/2}\Theta(v_z-v_{s}),
\label{jsin}
\end{equation}
where $v_p \equiv \Delta_p/v_f$, $v_s \equiv \Delta_{\rm {min}}/v_f$, and,
in three dimensions, $\rho \equiv 
\frac{1}{3}N_f v_f^2$.
The step function
reflects that the flow field $v_z$ must be sufficiently 
large, $v_z > v_{s}$, in order 
for nonlinear effects
to be present.

\subsubsection{3-D line quasinode}
For an energy gap as given
in (\ref{pcs}), where the nodal line is at $\theta = 
\pi/2$, $v_{fz}=0$ over the region of integration,
which is then limited to 
$|\theta-\pi/2|<\theta_c$.
Here
$(\theta_c-\pi/2)^2=\left[ ({\bf v}_f \cdot {\bf v})^2-\Delta_{\rm {min}}^2\right]/\Delta_p^2 
=
\left[ (v_f v_{\perp}\cos \eta)^2-\Delta_{\rm {min}}^2\right]/\Delta_p^2$,
where $v_{\perp}= (v_x^2+v_y^2)^{1/2}$ is the projection of ${\bf v}$ on the $x-y$ plane, and 
$\eta$ the angle between $v_{\perp}$ and the in-plane $v_f$.  
In our geometry $v_y=0$ and the only component of the nonlinear contribution
to the current is along $x$. 
We have,
\begin{equation}
j_{nl,x}= \frac{e N_f v_f \Delta_p}{2 \pi}\int^{\varphi_2}_{\varphi_1}
\int^{\theta_c}_{-\theta_c}  d\varphi d\theta
v_f \cos\varphi(\theta_c^2-\theta^2)^{1/2}\Theta(v_{\perp}-v_{s}).
\label{above}
\end{equation}
After performing the integration over $\theta$, this leaves an integral over
$\varphi$. To find the specific limits in this integral, we transform the 
integral over $\varphi$ to one over $\eta$. Using the relation  
$\varphi = \beta + \eta$, where $\beta$ is the (fixed) angle 
$v_{\perp}$ makes with the $x$ axis, we find,
\begin{equation}
j_{nl,x} = \frac{e N_f v_f }{4\Delta_p}\int^{\varphi_s}_{-\varphi_s} d\eta \cos(\beta+\eta)
[(v_f v_{\perp}\cos \eta)^2-
\Delta_{\rm {min}}^2]\Theta(v_{\perp}-v_{s}),
\label{ffff}
\end{equation} 
where $\varphi_s=\arccos\left(-\Delta_{\rm {min}}/v_f v_{\perp}\right)$. 
Making use of $\cos\beta = v_x/v_{\perp}$,
we get the nonlinear 
contribution to the current:
\begin{equation}
j_{nl,x} = \frac{e \rho}{v_p v_{\perp}^2} v_x 
\left(v_{\perp}^2-v_{s}^2\right)^{3/2}\Theta(v_{\perp}-v_{s}).
\label{explicitjc}
\end{equation}
 
\section {Perturbation solution}
\label{appb}

\subsection{2-D YBCO-type orthorhombicity}

Here we assume that the anisotropy factor, ${\Lambda_a}>1$, and 
first examine the case when $\psi \in [\psi_1,\psi_2]$, where 
the limiting angles $\psi_i$ are defined in the text. We
will need their expressions in the 
slab limit, $d \gg \lambda$, which are:
\begin{mathletters}
\label{xangles}
\begin{eqnarray}
\psi_1 &=& \tan^{-1}\left(\frac{\tan\alpha \,e^{\zeta_{s,a}(1-\Lambda_a)}}{\Lambda_a}\right), \\
\psi_2 &=& \tan^{-1}\left(\frac{\tan \alpha}{\Lambda_a }\right).
\end{eqnarray}
\end{mathletters}
However, unless otherwise stated, all expressions below are for arbitrary 
$d$.
Without loss of
generality, we give details of the solution
for the $a$-component of the nonlinear current, and
simply give the results for the $b$-component later, since it follows from 
an identical procedure.
We find after inserting (\ref{acurrent}) into (\ref{ML}),
\begin{mathletters}
\begin{eqnarray}
\label{orthode1}
\frac{d^2 u_a}{d \zeta_a^2}-{u_a}-2\varepsilon
u_a u_b\,\cos\alpha \sin\alpha&=&0, \qquad \zeta_a \in [\zeta_a^*,\zeta_{s,a}] \\
\frac{d^2 u_a}{d \zeta_a^2}-{u_a}+ \varepsilon[u_a^2\,
\cos^2\alpha  + u_b^2\,\sin^2\alpha]&=&0, \qquad \zeta_a \in [0,\zeta_a^*]
\label{orthode2}
\end{eqnarray}
\end{mathletters}
Here $\varepsilon = \Lambda_{a,n}^2 (H_A/H_0)\cos\alpha$,  
$\,\Lambda_{i,n}= \lambda_i/\lambda_n$ (for $i = a,b$ ), 
$\lambda_n^{-2} \equiv \frac{2 \pi e^2}{c^2} N_f v_f^2$.
We can now solve Eqs.~(\ref{orthode1}) and (\ref{orthode2}) perturbatively in the small
parameter $\varepsilon$, and write $u_i = u_{0,i} + \varepsilon u_{1,i}$. To zeroth order, we have
$u_{0,a} = \Lambda_a^{1/2}\sin\psi\,{\mathrm{sech}}(\zeta_{s,a}) \sinh(\zeta_a)$, and 
$u_{0,b} = -\Lambda_b^{1/2}\cos\psi\,{\mathrm{sech}}(\zeta_{s,b}) 
\sinh(\zeta_b)$, where $\Lambda_b \equiv \lambda_b/\lambda_a$. 
The first order solutions satisfy the following two equations:
\begin{mathletters}
\begin{eqnarray}
\label{orthode11}
\frac{d^2 u_{1,a}}{d \zeta_a^2}-{u_{1,a}}-2\,
u_{0,a} u_{0,b}\,\cos\alpha \sin\alpha&=&0,
\qquad \zeta_a \in [\zeta_a^*,\zeta_{s,a}] \\
\frac{d^2 u_{1,a}}{d \zeta_a^2}-{u_{1,a}}+ [u_{0,a}^2\,
\cos^2\alpha  + u_{0,b}^2\,\sin^2\alpha]&=&0, \qquad \zeta_a \in [0,\zeta_a^*].
\label{orthode22}
\end{eqnarray}
\end{mathletters}
The  boundary condition on the nonlinear terms is
$\partial{u_{1i}}/\partial{\zeta_i}|_{\zeta_{s,i}} = 0$.
By requiring continuity of the flow field, current, and magnetic field at the point $\zeta_a^*$,
we can obtain the first order solution $u_{nl,a} = \varepsilon u_{1,a}$ to
(\ref{orthode11}):
\begin{equation}
\label{unla}
u_{nl,a} = \Lambda_{a,n}^2 \cos\alpha\frac{H_A}{H_0}\Bigl[C_{1a} \cosh(\zeta_a) + C_{2a}\sinh(\zeta_a) + 
w_a(\zeta_a)\cosh(\zeta_a) + 
g_a(\zeta_a)\sinh(\zeta_a)\Bigr].
\end{equation}
Here the constants $C_{1i}$ and $C_{2i}$ 
are given by $C_{1i} = \overline{w}_i(\zeta_i^*)-\overline{w}_i(0)
-w_i(\zeta_i^*)$, $C_{2i} = -g_i(\zeta_{s,i})+\tanh(\zeta_{s,i})[\overline{w}_i(0)+w_i(\zeta_i^*)-
w_i(\zeta_{s,i})-
\overline{w}_i(\zeta_i^*)]$, with $\overline{w}_a\equiv -\mu_2{{U}}_{a}-\mu_3{{W}}_{c,a}$, 
$w_a \equiv -\mu_1{{G}}_{c,a}$, $g_a \equiv \mu_1 {{G}}_{s,a}$,
and
\begin{mathletters}
\begin{eqnarray}
{{G}}_{s,i}(\zeta_i) &=&\frac{\sinh(({\Lambda_i}+2)\zeta_i)}{4({\Lambda_i}+2)} - 
\frac{\sinh(({\Lambda_i}-2)\zeta_i)}{4({\Lambda_i}-2)}, \\
{{G}}_{c,i}(\zeta_i) &=&\frac{\cosh((\Lambda_i + 2)\zeta_i)}{4(\Lambda_i + 2)} + 
\frac{\cosh(({\Lambda_i}-2)\zeta_i)}
{4({\Lambda_i}-2)}-\frac{\cosh( {\Lambda_i}\,\zeta_i)}{{2\Lambda_i}}, \\
{{W}}_{c,i}(\zeta_i) &=&\frac{\cosh((2{\Lambda_i}+1)\zeta_i)}{4({2\Lambda_i}+1)}
-\frac{\cosh((2{\Lambda_i}-1)\zeta_i)}{4({2\Lambda_i}-1)}-\frac{\cosh(\zeta_i)}{2}, \\
{{U}}_{i}(\zeta_i) &=&\frac{1}{12}\cosh(3\,\zeta_i)-\frac{3}{4}\cosh(\zeta_i), \\
\mu_1 &=& -\frac{\sin 2 \alpha\sin2\psi}{2\cosh(\zeta_{s,a})
\cosh(\zeta_{s,b})}, \\
\mu_2 &=&-\frac{\Lambda_a\cos^2\alpha \sin^2 \psi}{\cosh^2(\zeta_{s,a})}, \\
\mu_3 &=& -\frac{\Lambda_b\sin^2\alpha\sin^2\psi}{\cosh^2(\,\zeta_{s,b})}.
\end{eqnarray}
\end{mathletters}
Similarly, (\ref{orthode22}) gives
\begin{equation}
\label{unlabar}
{u}_{1,a} = C_{3a} \cosh(\zeta_a) + C_{4a}\sinh(\zeta_a) + \overline{w}_a(\zeta_a)\cosh(\zeta_a) + 
\overline{g}_a(\zeta_a)\sinh(\zeta_a),
\end{equation}
where $C_{3i} = -\overline{w}_i(0), C_{4i} = 
C_{2i}+g_i(\zeta_i^*)-\overline{g}_i(\zeta_i^*)$. 
Here $\overline{g}_a \equiv \mu_2 {V_a}+
\mu_3 W_{s,a}$, and
\begin{eqnarray}
{W_{s,i}}(\zeta_i) &=&\frac{\sinh(( 
2{\Lambda_i}-1)\zeta_i)}{4({2\Lambda_i}-1)} +
\frac{\sinh((2{\Lambda_i}+1)\zeta_i)}{4(2{\Lambda_i}+1)}-\frac{\sinh(\zeta_i)}{2}, \\
{V_{i}}(\zeta_i) &=&\frac{1}{4}\sinh(\zeta_i)-\frac{1}{12}\sinh(3\zeta_i).
\end{eqnarray}
The
matching point $\zeta_a^*$  (for $\zeta_{s,a} \gg 1$) is :
\begin{equation}
\zeta_a^* = \zeta_{s,a}-
\frac{1}{1-{\Lambda_a}}\ln\left(\frac{\Lambda_a\tan\psi}{\tan\alpha}\right).
\end{equation}
The $b$-component is found by a similar procedure. We find, 
\begin{eqnarray}
\label{unlb}
u_{nl,b} &=& \Lambda_{b,n}^2 \sin\alpha\frac{H_A}{H_0}\Bigl[C_{1b} \cosh(\zeta_b) + 
C_{2b}\sinh(\zeta_b) + w_b(\zeta_b)\cosh(\zeta_b) + 
g_b(\zeta_b)\sinh(\zeta_b)\Bigr], \\ \nonumber
 \qquad \zeta_b &\in& [\zeta_b^*,\zeta_{s,b}],\\
u_{nl,b} &=& \Lambda_{b,n}^2 \sin\alpha\frac{H_A}{H_0} \Bigl[C_{3b} \cosh(\zeta_b) + 
C_{4b}\sinh(\zeta_b) + \overline{w}_b(\zeta_b)\cosh(\zeta_b) + 
\overline{g}_b(\zeta_b)\sinh(\zeta_b)\Bigr], \\ \nonumber
\qquad \zeta_b &\in& [0,\zeta_b^*],
\label{unlbbar}
\end{eqnarray}
where, $w_b = \mu_2 W_{c,b} + \mu_3 U_b$, $g_b = -\mu_2 W_{s,b}-\mu_3 V_b$, 
$\overline{w}_b = \mu_1 G_{c,b}$, and
$\overline{g}_b = -\mu_1 G_{s,b}$.

The fields calculated above are  for $\psi \in [\psi_1,\psi_2]$.
For $\psi \in [0,\psi_1]$, we get $u_{nl,i}$ by simply setting the crossover point $\zeta_i^* = 0$ 
in Eqs.~ (\ref{unla},\ref{unlb}). Similarly, to find the nonlinear fields for $\psi \in [\psi_2,\pi/2]$,
we set $\zeta_i^* = \zeta_{s,i}$ in Eqs.~(\ref{unlabar},\ref{unlbbar}).

\subsection{3-D quasi-nodes}

\subsubsection{3-D point quasinodes}

For the geometry we
consider, ${\bf H}_{\rm{A}}$ is in the $a-c$ plane, and  
the nonlinear fields now have only a $z$-component, which depends    
on the coordinate $y$. The flow field decreases rapidly with distance into the sample, so that there
will be a point in the material, $\zeta_z^*$, where $u_z<\kappa$, so that
nonlinear corrections are absent for distances below $\zeta_z^*$.
Again, there is no restriction on sample thickness, unless otherwise stated.
Inserting the current (\ref{jsin}) into (\ref{ML}) gives:
\begin{mathletters}
\begin{eqnarray} 
\label{devst} 
\frac{d^2 u_z}{d \zeta_z^2}-{u_z}+{\varepsilon}
\left(u_z^2-\kappa^2\right)^{3/2}\Theta(u_z-\kappa)=0,
\end{eqnarray} 
\end{mathletters}
where 
$\varepsilon \equiv h^2=(H_A/H_0)^2$, and $\kappa$ and $H_0$ 
are defined in the text.
(\ref{devst}) can be written as
\begin{mathletters}
\begin{eqnarray} 
\label{devs} 
\frac{d^2 u_z}{d \zeta_z^2}-{u_z}+{\varepsilon}
(u_z^2-\kappa^2)^{3/2}&=&0, \qquad \zeta_z\in[\zeta_z^*,\zeta_{s,z}], \\
\frac{d^2 {u}_z}{d \zeta_z^2}-{{u}_z}&=&0,\qquad \zeta_z\in[0,\zeta_z^*].
\label{devss}
\end{eqnarray} 
\end{mathletters}
We now solve (\ref{devs}) perturbatively to first order, and write 
$u_{nl,z}= \varepsilon \,u_{1z}$.
We find,
\begin{equation}
\label{3dsol}
u_{nl,z}(\zeta_z) = \frac{H_A^2}{H_0^2}[D_1\cosh(\zeta_z) + D_2\sinh(\zeta_z) 
+k(\zeta_z)\cosh(\zeta_z)+f(\zeta_z)\sinh(\zeta_z)],
\end{equation}
where the constants $D_1$ and $D_2$ are 
found by requiring continuity of the fields at $\zeta_z^*$, and
given by $D_1 = -k(\zeta_z^*)$, 
$D_2=-f(\zeta_{s,z})+\tanh(\zeta_{s,z})[k(\zeta_z^*)-k(\zeta_{s,z})]$.
The functions $f$ and $k$ are found by elementary methods, and are given by,
\begin{mathletters}
\begin{eqnarray}
\label{3dpointg}
f(\zeta_z)&=& \frac{1}{8}\biggl(-\frac{3 \kappa^4}{ m(\psi)}\ln 
\biggl[2\left(m(\psi)\sinh(\zeta_z)+\sqrt{m^2(\psi)\sinh^2(\zeta_z)-\kappa^2}\right)\biggl] \\ \nonumber
&+&\left[5\kappa^2\sinh(\zeta_z)-2 m^2(\psi)\sinh^3(\zeta_z)\right]
\left[m^2(\psi)\sinh^2(\zeta_z)-
\kappa^2\right]^{1/2}\biggr)\Theta\Bigl(m(\psi)\sinh(\zeta_z)-\kappa\Bigr),\\
\label{3dpointw}
k(\zeta_z)&=& \frac{1}{8}\biggl(
\frac{3}{m(\psi)} (m^2(\psi)+\kappa^2)^2\,
\ln\biggl[\sqrt{2}\left(m(\psi)\cosh(\zeta_z)+\sqrt{m^2(\psi)\sinh^2(\zeta_z)-\kappa^2}\right)\biggl]
\\ \nonumber
&+&\cosh(\zeta_z)\left[-4 m^2(\psi)-5\kappa^2+m^2(\psi)\cosh(2\zeta_z)\right] \\ \nonumber
&\times&
\left[m^2(\psi)\sinh^2(\zeta_z)-\kappa^2\right]^{1/2}\biggr)
\Theta\Bigl(m(\psi)\sinh(\zeta_z)-\kappa\Bigr),
\end{eqnarray}
\end{mathletters}
where $m(\psi) \equiv \sin\psi/\cosh\zeta_{s,z}$.
The matching point is found to be
$\zeta_z^* = \sinh^{-1}[\kappa/m(\psi)]$.
In a similar fashion, (\ref{devss}) has the first order solution 
\begin{equation}
{u}_{1z} =  \left[f(\zeta_z^*)-f(\zeta_{s,z})+
\tanh(\zeta_{s,z})(k(\zeta_z^*)-k(\zeta_{s,z}))\right]\sinh(\zeta_z).
\end{equation}

\subsubsection{3-D line quasinode.}
${\bf H}_{\rm{A}}$ is again in the $a-c$ plane, and due to the form of the gap,  
the nonlinear fields now have only a $x$-component, which depends    
on the coordinate $y$.
Here $u_x>\kappa$, in order for
nonlinear effects to be present. 
These effects are therefore absent at distances below $\zeta^*$.
Inserting (\ref{explicitjc})
into (\ref{ML}) gives the following:
\begin{eqnarray} 
\label{devc} 
\frac{d^2 u_x}{d \zeta^2}-{u_x}+\frac{\varepsilon}{u_x}\left(u_x^2-
\kappa^2\right)^{3/2}\Theta(u_x - \kappa)&=&0, 
\label{devcc}
\end{eqnarray} 
where the small parameter $\varepsilon$ is 
$\varepsilon \equiv h \equiv H_A/H_0$.

We can now solve (\ref{devc}) perturbatively to first order. 
We write $u_{x} = u_{0x} + \varepsilon u_{1x}$.
To avoid unnecessarily tedious calculations,
we take the slab limit, and hence the zeroth
order solution, $u_{0x}$ is,
$u_{0x} \approx-\cos\psi \exp{(\zeta-\zeta_s)}$.
The first order solution, $u_{1x}$, is found from dividing the slab at $\zeta^*$ into
two portions. This allows (\ref{devc}) to be written
\begin{mathletters}
\begin{eqnarray} 
\label{devc1} 
\frac{d^2 u_x}{d \zeta^2}-{u_x}+\frac{\varepsilon}{u_x}\left(u_x^2-
\kappa^2\right)^{3/2}&=&0, \qquad \zeta\in[\zeta^*,\zeta_{s}],\\
\frac{d^2 u_x}{d \zeta^2}-{u_x}&=&0,\qquad \zeta\in[0,\zeta^*].
\label{devcc2}
\end{eqnarray} 
\end{mathletters}
The
solution to (\ref{devc1}) is found by methods similar to the point node case, and is given by
\begin{eqnarray}
u_{1x} = E_1\,e^{\zeta} + E_2\,e^{-\zeta} + r(\zeta)\,e^{\zeta} + s(\zeta)\,e^{-\zeta},
\end{eqnarray}
The functions $r(\zeta)$
and $s(\zeta)$ are found by elementary methods, and
the constants $E_1$ and $E_2$ are determined from
 the boundary conditions
$(\partial{u_{1x}}/\partial{\zeta})|_{\zeta_s} = 0$,  and 
continuity of the flow field, current, and magnetic field at the point $\zeta^*$:
\begin{mathletters}
\begin{eqnarray}
E_1 &=&  {(s(\zeta_s)-s(\zeta^*)-r(\zeta^*))e^{-2\zeta_s}+ 
r(\zeta_s)}\\
E_2 &=& {-s(\zeta_s)e^{-2\,\zeta_s}+r(\zeta_s)-s(\zeta^*)- r(\zeta^*)}
\\
r(\zeta)&=& \frac{1}{8}\biggl[-3n(\psi)\kappa\tan^{-1}\left(\frac{n^2(\psi)\,e^{2\,\zeta}-
2\kappa^2}{2\kappa\,\sqrt{n^2(\psi)\,e^{2\,\zeta}-\kappa^2}}\right)\\ \nonumber
&+&\frac{2}{n(\psi)}
(2\,n^2(\psi)+\kappa^2 e^{-2\zeta})\sqrt{n^2(\psi) e^{2\zeta}-
\kappa^2}\biggr]\Theta\left(n(\psi)e^{\zeta}-\kappa\right),\\
s(\zeta)&=&  \frac{1}{6}\biggl[-\frac{3\kappa^3}{n(\psi)}\tan^{-1}
\left(\frac{\sqrt{n^2(\psi)\,e^{2\zeta}-\kappa^2}}
{\kappa}\right) \\ \nonumber 
&+&\frac{1}{n(\psi)}(4\kappa^2-n^2(\psi)\,e^{2\zeta})\sqrt{n^2(\psi) e^{2\zeta}-
\kappa^2}\biggr]\Theta\left(n(\psi)e^{\zeta}-\kappa\right),
\end{eqnarray}
\end{mathletters}
where $n(\psi) \equiv \cos\psi\exp{(-\zeta_s)}$.
Similarly, (\ref{devcc2}) can be solved, with the only major 
difference begin the boundary condition
$u_{1x}(0) = 0$.
We find,
\begin{equation}
{u}_{1x} =  E_3 \left(e^{\zeta}-e^{-\zeta}\right),
\end{equation}
where
\begin{equation}
E_3 = {(s(\zeta_s)-s(\zeta^*))e^{-2\,\zeta_s}-r(\zeta_s) +
r(\zeta^*)}.
\end{equation}
The matching point $\zeta^*$
is found by equating (\ref{devc1}) and (\ref{devcc2}) 
at $\zeta^*$, giving
$\zeta^* = \zeta_s+ \ln[\kappa/\cos\psi]$.

%
%

\begin{figure}
\caption{Variables and labels for the OP in 
subsection \protect{\ref{sybco}}. The four nodal lines are labeled by the
numbers (1) through (4). 
The Fermi velocity at  node (1), ${\bf{v}}_f$, forms
an angle $\alpha$ with the $a$-axis. The applied field
${\bf{H}}_{\rm{A}}$ is at an angle $\psi$ to the $a$-axis. A generic
flow field vector $\bf{v}$ is shown
for illustrative purposes.}
\label{FS}
\end{figure}

\begin{figure}
\caption{Angular dependence of $\Delta \lambda$ for
the OP in subsection \ref{sybco}, with YBCO-type orthorhombicity. We
plot here the function ${{\cal Y}(\psi)}$ 
(Eqs.~\protect{\ref{ybcowhole}-\ref{ybcowhole3}}), vs. the angle $\psi$. 
This function
represents the angular dependence of $\Delta \lambda$ normalized
by the field dependence and numerical prefactors in (\ref{calp}).
In this Figure, the Fermi velocity direction is fixed at an angle 
$\alpha =\pi/4$. 
The bolder line shows the result for  
the orthorhombicity parameter $\Lambda_a \equiv \lambda_a/\lambda_b=1.0$
(tetragonal limit), while $\Lambda_a$ 
equals  1.1, 1.2, 1.3, 1.4 and 1.5 in the other curves (from top to
bottom at $\psi=0$).}
\label{varyL}
\end{figure}

\begin{figure}
\caption{Angular dependence of $\Delta \lambda$, 
for the same OP as
in Fig.~\ref{varyL}. Again, the function ${{\cal Y}(\psi)}$ is plotted.
Here $\Lambda_a = 1.0$ is fixed, while $\alpha$ takes the values
$\pi/4 \pm n(\pi/80)$ with $n=0, 1, 2, 3$. As in the previous Figure, 
the effect of orthorhombicity is quite strong. The bold line is the
tetragonal ($n=0$) case. The positive values of $n$ are (dotted
curves) above the
bold curve at small angles, and those for negative $n$ (dashed)
are below.}
\label{alpha}
\end{figure}

\begin{figure}
\caption{Angular dependence of $\Delta \lambda$ 
for the same OP as
in Figs. \ref{varyL} and \ref{alpha}. 
Here $\Lambda_a = 1.3$ is fixed, while $\alpha$ takes the same values
as in the previous Figure. The solid curve corresponds
to $\alpha=\pi/4$ and the meaning of the dotted and dashed lines is
as in the previous Figure. The combined effects of the two anisotropy 
parameters are seen.}
\label{both}
\end{figure}

\begin{figure}
\caption{Angular dependence of $\Delta \lambda$ for a 2-D OP
with line nodes
and BSCCO-type orthorhombicity (subsection \ref{sbssco}). 
The applied field makes an angle $\psi$ with the new
orthorhombic $a$-axis. The nodal lines are now along the principal axes, 
which are 
rotated by an angle of $\pi/4$ relative
to the previous undistorted tetragonal axes.
The curves are normalized so that they
represent the function ${\cal B}(\psi)$ as defined
in (\ref{2dbssco}). 
They correspond, from top to bottom at
$\psi=0$ to values of $\Lambda_a = 1.0, 1.1, 1.2, 1.3, 1.4 , 1.5$. As in
previous Figures, the tetragonal limit is plotted with
a bolder line. }
\label{bsscofig}
\end{figure}

\begin{figure}
\caption{Angular dependence of $\Delta \lambda$ 
for a 2-D gap with quasinodes and tetragonal symmetry
as discussed in subsection \ref{s2dqn}. Results are normalized so
that the quantity plotted is the function ${\cal Q}$ of (\ref{Q}).
This function is plotted vs. the angle $\psi$ that the applied field forms
with the $a$ axis. The bolder line corresponds to the usual result
with nodes, ($\kappa=0$ see (\ref{kappaeq})), and the other curves, from top to
bottom, are for values of $\kappa$ of 0.2, 0.4, 0.6 and 0.8 respectively.} 
\label{quasi2d}
\end{figure}

\begin{figure}
\caption{Field dependence of $\Delta \lambda(\psi)$ for the same
2-D gap with line quasinodes considered in the previous Figure. 
The curves show the quantity $\Delta \lambda/\lambda$ (see \ref{2dqn})
at $\psi=\pi/4$,
as a function of $h$ for values of $\delta$ (see text) equal to 
0, 0.01, 0.02, 0.03, 0.04 and 0.05 These values can be read off from the
threshold field values in the curves.}
\label{quasi2dh}
\end{figure}

\begin{figure}
\caption{Angular and field dependence of $\Delta \lambda(\psi)$ for a 
3-D gap with point quasinodes. 
The main plot is normalized so that the quantity plotted is
the function ${\cal P}(\psi,\kappa)$ defined in (\ref{lampoints}).
This quantity is plotted as a function of angle for
values of $\kappa = 0,0.1,0.2,0.3,0.4,0.6,0.8$, as defined
in (\ref{kappa2}). The bolder
curve is the $\kappa=0$ result. The inset displays the field dependence 
of $\Delta\lambda$ at $\psi=\pi/2$. The quantity plotted is $\Delta \lambda/\lambda_z$ vs.
normalized applied field, for values of $\delta$ equal to
0, 0.01, 0.02, 0.03, 0.04 and 0.05.}
\label{3dpoints}
\end{figure}

\begin{figure}
\caption{Angular dependence of $\Delta \lambda(\psi)$ for a 
3-D line quasinode. 
The main plot is normalized so that the quantity plotted is
the function ${\cal L}(\psi,\kappa)$ defined in (\ref{lamline}).
The inset has the field dependence 
of $\Delta\lambda$ at $\psi=0$.
All parameter values are precisely as
in the previous Figure.}
\label{3dline}
\end{figure}

%
%

\begin{references}
\bibitem{old} In Ref.{\onlinecite{zv2}}, the customary  
 distinction between the presence of an $s$ wave gap
and FS anisotropy was kept. We have here eliminated this distinction
and recalculated the fields in terms of the best minimum set
of variables.
\bibitem{phot} The angles $\phi_1$ can be inferred from photoemission
or other experiments.
\end{references}
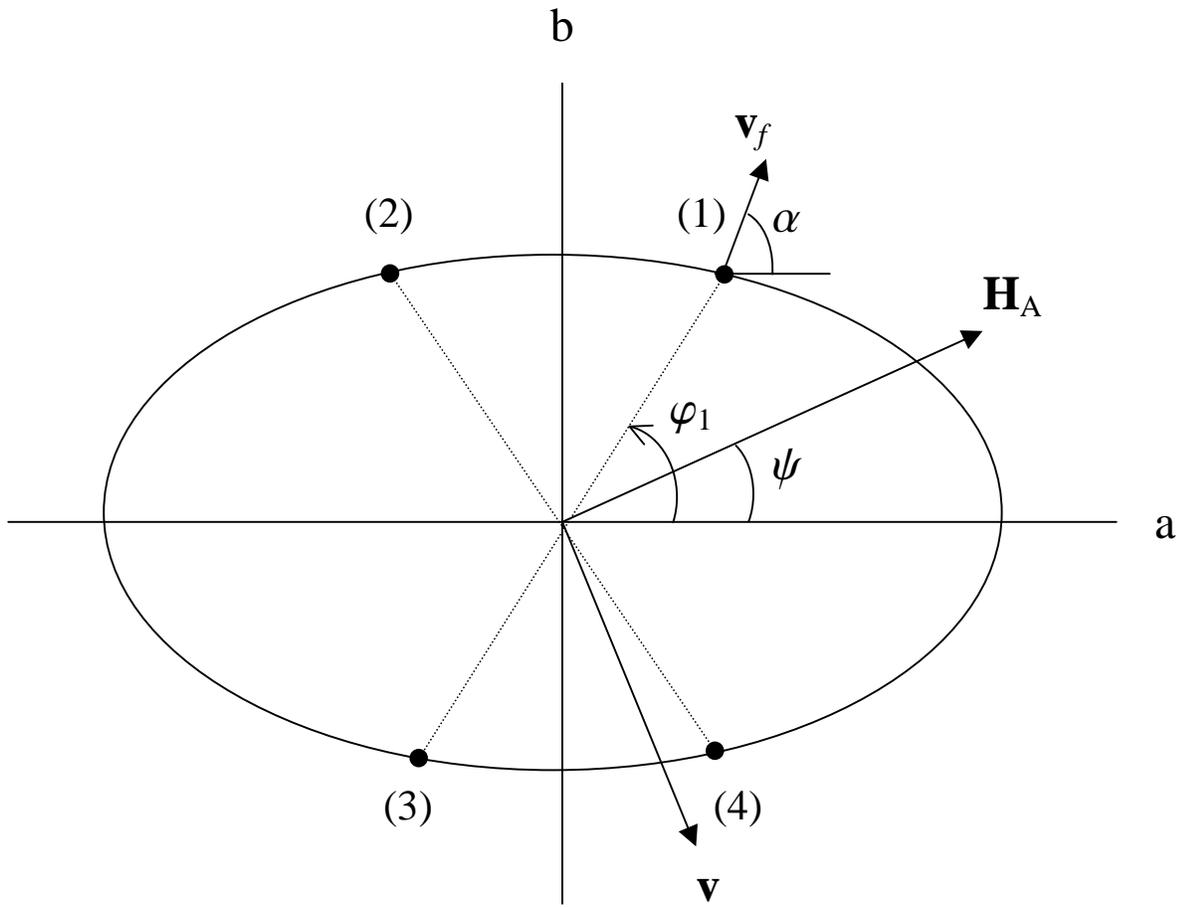
\end{document}